\pdfoutput=1

\documentclass[aps, prl, twocolumn]{revtex4-1}

\usepackage{amsmath}
\usepackage{amssymb}
\usepackage{color}

\usepackage{hyperref}

\def\flushboth{%
  \let\\\@normalcr
  \@rightskip\z@skip \rightskip\@rightskip
  \leftskip\z@skip
  \parindent 1.5em\relax}

\usepackage{verbatim}
\usepackage{graphicx}
\usepackage{subfig}

\newcommand{\bi}{\begin{itemize}}
\newcommand{\ei}{\end{itemize}}
\newcommand{\be}{\begin{equation}}
\newcommand{\ee}{\end{equation}}

\newcommand{\bml}{\begin{multline}}
\newcommand{\emll}{\end{multline}}

\def\({\left(} \def\){\right)}
\def\[{\left[} \def\]{\right]}

\def\v{\vec}

\newcommand{\G}{\Gamma}

\newcommand{\la}{\langle}
\newcommand{\ra}{\rangle}

\newcommand{\bea}{\begin{eqnarray}}
\newcommand{\eea}{\end{eqnarray}}

\def\ie{\begin{equation}\begin{aligned}}
\def\fe{\end{aligned}\end{equation}}

\def\1{{\mathds 1}}

\usepackage[UKenglish]{datetime}

\begin{document}

\title{Chaos in  the Quantum Field Theory $S$-matrix}

\author{Vladimir Rosenhaus\\[-5pt] \text{}}
\affiliation{\\ School of Natural Sciences, Institute for Advanced Study, \\  Einstein Drive, Princeton, NJ 08540\\ \& \\ 
Initiative for the Theoretical Sciences, The Graduate Center, CUNY, 
New York, NY 10016}


\begin{abstract}
A number of studies have shown that chaos occurs in scattering: the outgoing deflection angle is seen to be an erratic function of the impact parameter. We propose to extend this to quantum field theory, and to use  erratic behavior of the many-particle $S$-matrix as a probe of chaos.
\end{abstract}
\maketitle

\maketitle

\section{Introduction}

How can we characterize chaos in quantum field theory? Based on intuition from studies of chaotic scattering in  classical  and quantum mechanics, we will propose that chaos is visible in the quantum field theory $S$-matrix.

Chaos in  classical physics is characterized by stretching and folding: a region of phase space experiences  stretching along unstable directions, with an average rate given by the Lyapunov exponents, along  with folding, in order to remain confined to a finite region of phase space. An initially localized patch of phase space evolves into a highly  complex structure which is spread throughout the available phase space, while still maintaining the same volume as the initial patch, as required by Liouville's theorem.

In quantum systems, the state space is not phase space, but rather that of the eigenfunctions and eigenvalues of the Hamiltonian. Quantum systems which are classically chaotic are known to exhibit universal features, such as Wigner-Dyson statistics for energy eigenvalues \cite{BGS}.  Semiclassical theory based on the Gutzwiller trace formula \cite{Gutzwiller70}, a kind of generalization of Bohr-Sommerfeld quantization, provides a bridge between the spectral data of the quantum Hamiltonian and the dynamics of the classical system.

 Chaos in many-body systems and field theories is far more challenging to study, yet is of essential importance, as it provides the microscopic underpinning of thermodynamics \cite{FPU, ETH1, ETH2, ETH3}. A recent insight has been to extend  the application of the out-of-time-order correlator \cite{LO} - which gives a quantum mechanical analog of a  Lyapunov exponent - to quantum field theories with a large number of fields \cite{Kitaev, MSS}, such as the SYK model \cite{SY, KitaevSYK, PR, Maldacena:2016hyu, Rosenhaus:2019mfr}. 
However, a deeper understanding of chaos in quantum field theory will require knowing much more. Certainly within classical mechanics, knowing that there are positive Lyapunov exponents does not, by itself, tell one of the full richness of chaotic dynamics.

\begin{figure}[t]
\centering 
\subfloat[]{
\includegraphics[width=1.3in]{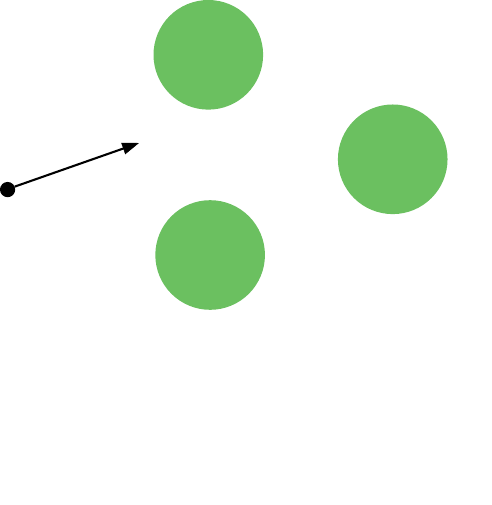}  
} \\[10pt] 
\subfloat[]{
\includegraphics[width=3.4in]{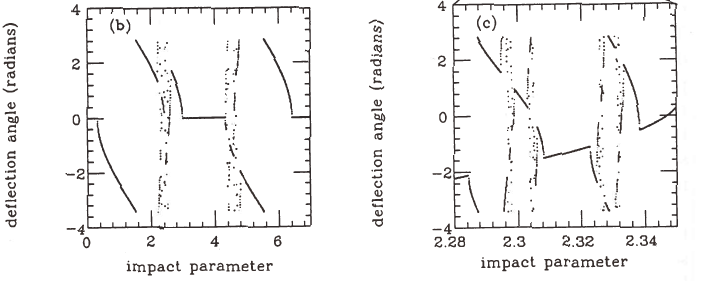}
}
\caption{  (a) A particle  scattering off of three disks. (b) A plot from \cite{Smilansky} of the outgoing deflection angle as a function of the impact parameter. The left plot includes the full range of impact parameters, while the right plot zooms in on a subset. The plot clearly demonstrates chaos: the deflection angle undergoes wild fluctuation under variation of the impact parameter. } \label{ThreeBalls}
\end{figure}

In looking for a characterization of chaos in quantum field theory, we  will neither seek a quantum generalization of chaos in classical field theory, nor shall we seek  the continuum limit of a lattice system, either classical or quantum. Instead, we will  immediately go to the natural observable of quantum field theory: the $S$-matrix. In a scattering experiment, one starts with far-separated wavepackets at early times, which interact at intermediate times, leading to far-separated wavepackets at late times. The $S$-matrix is the overlap of the $in$ state with the $out$ state. The vast majority  of the infinite number of degrees of freedom that a field theory possesses, and which make it so challenging to study chaos, are never excited in a scattering experiment.

The phenomenon of chaos in scattering, while  less familiar than chaos in systems with bounded phase space, is well-established.
We will recall a few  canonical examples in classical and quantum mechanics, and then we will propose this be extended to the  $S$-matrix in quantum field theory. A special case, which has recently been discussed, is chaos in the black hole $S$-matrix \cite{Polchinski}.

\section{Chaos in scattering} \label{Sec2}

A classic example of chaotic scattering, discussed in \cite{Smilansky}, is that of elastic scattering of a particle against three fixed disks \cite{Eckhardt1987, Gaspard1, Gaspard2, Gaspard3,  Cvitanovich, ChaosBook}, as shown in Fig.~\ref{ThreeBalls} (a). The particle enters with some impact parameter, it  scatters off of the three disks,  perhaps hitting several of the disks multiple times, and it emerges at some outgoing scattering angle. A plot of the scattering angle versus the impact parameter is shown in Fig.~\ref{ThreeBalls} (b). Strikingly, the scattering angle is a highly erratic function of the impact parameter: the system is chaotic.

The result appears puzzling at first: for most impact parameters, the particle will undergo only a few collisions with the disks before emerging, and restricting to those initial conditions would surely not give an erratic scattering angle. Indeed, the fraction of impact parameters that lead to the particle  spending longer than time $T$ within the scattering region decays exponentially, $e^{- \gamma T}$. However, crucially, there exist infinitely many impact parameters that lead to the particle spending an arbitrarily long amount of time within the scattering region, bouncing around between the disks. It is this property which gives the erratic behavior shown in Fig.~\ref{ThreeBalls} (b): an arbitrarily small range of impact parameters will lead the particle to scatter into the full range of deflection angles.

The quantum version of  a particle scattering off of a potential  is much richer. Semiclassically, one can form a wavepacket of high momentum modes and, for instance, study possible interference effects of classical paths. More generally, quantum scattering is described  by the $S$-matrix, which, in the basis of plane waves, is the overlap of the $in$ state and the $out$ state, $S(\v p \rightarrow \v q) = {}_{out}\la q| p\ra_{in}$, where $\v p$ and $\v q$ are the ingoing and outgoing momenta, respectively. One can alternatively consider a discrete basis of states, $|n\ra$,  so that the $S$-matrix is a matrix, $S_{nm}(E) = {}_{out} \la m| n\ra_{in}$, where $E$ is the energy of the state. 
 Extrapolating from  examples, Bl\"umel and Smilansky \cite{Blumel1, Blumel2, Blumel3}  conjectured that, for systems whose classical analog is chaotic, ``the statistical properties of the $S$-matrix (for $\hbar\rightarrow 0$) are determined by Dyson's theory for the orthogonal ensemble of random unitary matrices'' \cite{Smilansky}. 

\begin{figure}[t]
\centering 
\subfloat[]{
\includegraphics[width=1in]{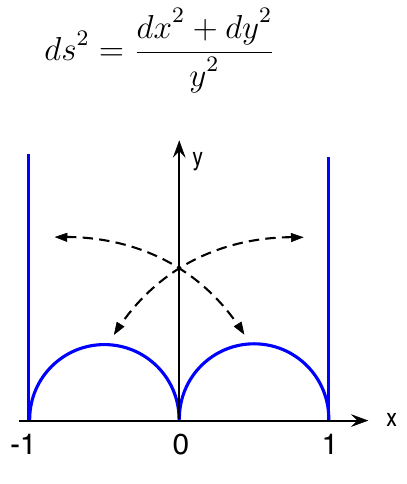}
}   \ \ \ \ \
\subfloat[]{
\includegraphics[width=1in]{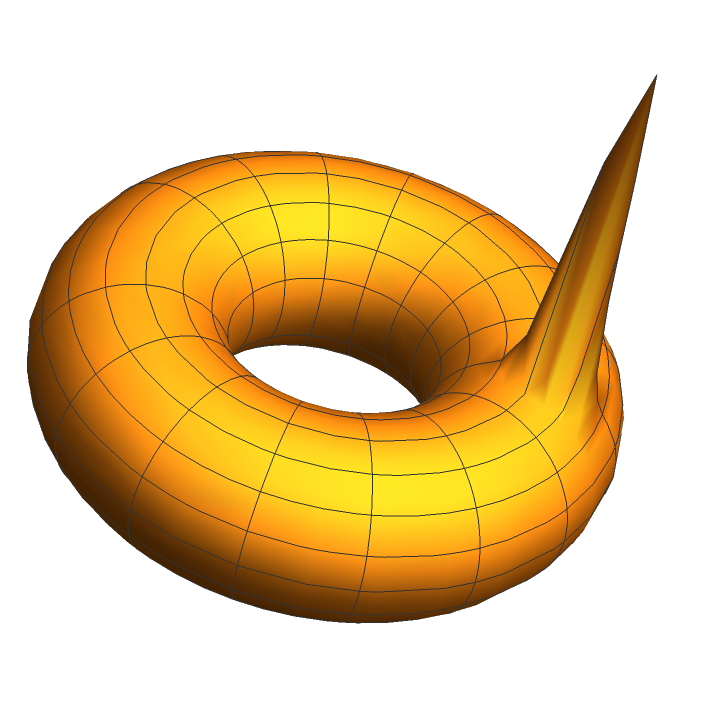}
}\ \ \ 
\subfloat[]{
\includegraphics[width=1.5in]{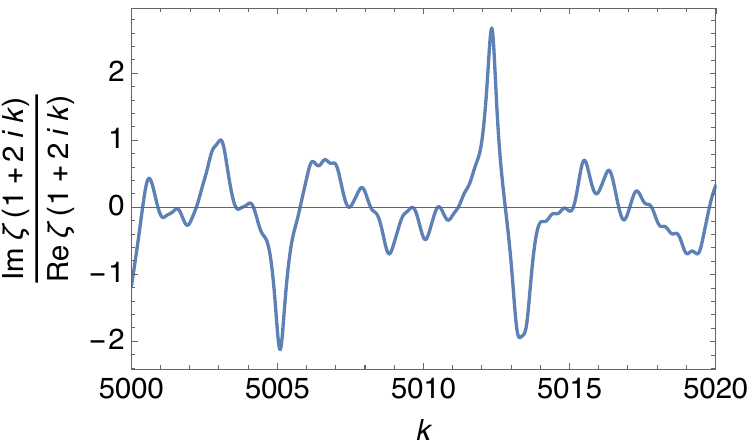}
}
\caption{   A piece of hyperbolic space is cut out and the sides are identified, as indicated in (a). The result is a torus with a cusp that extends to infinity (a ``leaky torus''), as shown in (b).  One considers a quantum mechanics scattering problem on the leaky torus. A wave is sent into the torus from the end of the cusp, at infinity. The relative phase of the reflected wave involves the Riemann zeta function: an analytic yet erratic function, as shown in (c).} \label{Horn}
\end{figure}

For most chaotic systems, the $S$-matrix must be found numerically. A remarkable analytic example is that of a leaky torus \cite{GutzwillerBook, Gutzwiller,   Lax, Faddev}: one cuts out a piece of hyperbolic space and identifies the sides, as shown in Fig.~\ref{Horn}. The result is a torus, of negative curvature, that has a cusp extending to infinity~\cite{Note1}. In terms of   Fig.~\ref{Horn} (a), $y=\infty$ along with the three points at $y=0$: $x=-1,0,1$, are identified with infinity. One can send ingoing waves from infinity through the cusp, and observe the phase shift of the outgoing waves.   The solution to the Schr\"odinger equation at large $y$,  and after integrating over the $x$ direction, is a superposition of an incoming and outgoing wave, $y^{\frac{1}{2}- i k} +S(k)\, y^{\frac{1}{2}+i k}$, with the $S$-matrix, 
\be
S(k) = \frac{Z(1+ 2 i k)}{Z(1 - 2 i k)} ~\  \  \ \text{where} \ \ \ \ Z(x) = \pi^{-x/2} \G(x/2) \zeta(x)~,
\ee
where $\zeta(x)$ is the Riemann zeta function, see Fig.~\ref{Horn} (c). The phase is erratic: even though the Riemann zeta function is analytic, it is seemingly unpredictable.

\section{Chaos in the quantum field theory $S$-matrix} \label{Sec3}

\begin{figure}[tbp]
\centering
\includegraphics[width=1in]{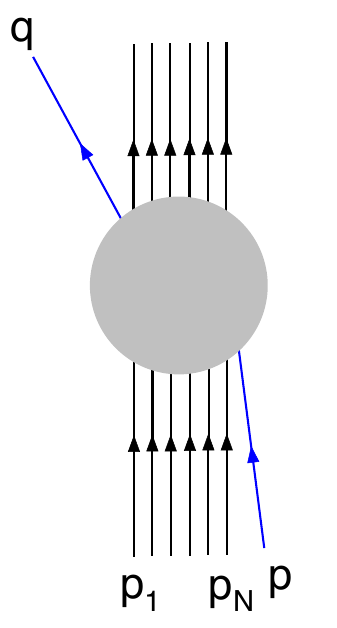}
\caption{We expect a many-particle $S$-matrix in quantum field theory to exhibit chaos. The interactions occur within the shaded region. In certain settings, one may view the collection of particles  $p_i$, with $i=1, \ldots, N$,  as analogous to the three disks in Fig.~\ref{ThreeBalls}. } \label{SQFT}
\end{figure}

We would like to generalize the discussion of chaos in nonrelativistic scattering to relativistic quantum mechanics. This requires the language of quantum field theory. In quantum field theory, one computes the $n$ to $m$ $S$-matrix, $
{}_{out}\la p_1',\ldots, p_m'| p_1,  \ldots, p_n\ra_{in}$, where the $in$ state in the asymptotic past consists of $n$ particles with $d$-dimensional momenta $p_i$, and the $out$ state in the asymptotic future consists of $m$ particles with momenta $p_i'$. \\[-20pt]
\subsubsection*{Proposal}
We propose that the $N{+}1$ to $M{+}1$ quantum field theory $S$-matrix, 
\be \label{prop}
{}_{out} \la p_1', \ldots, p_M'; q| p_1, \ldots, p_N; p\ra_{in}~,
\ee
in the limit  $N+M\gg 1$, may exhibit chaos in some regions of phase space, in the sense of erratic behavior under  variation of some of the individual momenta such as $p$. 
\\[1pt]

The motivation for this proposal comes from the example of the three-disk scattering problem discussed in the previous section. 
The quantum field theory $S$-matrix is much richer than the one-to-one $S$-matrix in the  mechanics problem of a particle scattering off of a potential. To imitate the latter, we think of the $N$ particles with momenta $p_i$ as playing the role of the potential (the three disks), and the additional particle with incoming momentum $p$  and outgoing momentum $q$ as playing the role of the particle scattering off of the potential. We must take $N$ to be large, in order to have a state that is far from the vacuum. Note that in the three-disk problem, a single disk is already a state that is far from the vacuum; for quantum field theory, each particle is a fundamental excitation, so we need a state with many particles \cite{Note3}.
 
A possibly instructive setup to study, which models a static potential, is in a theory with two species of fields, one heavy and one light. The light field is taken to have an interaction  with the heavy field, but no self-interaction. Taking the $N$ particles with momentum $p_i$ in the $in$ state to be excitations of the heavy field, and the $in$ particle with momentum $p$ to be an excitation of the light field, the $S$-matrix will be concentrated around $N\approx M$ and $p_{i}' \approx p_i$ (since $N$ is large, $q$ can  differ significantly from $p$, with the process still conserving momentum). 
The incoming particle with momentum $p$, after interacting with the $N$ other particles, will exit with momentum $q$. 
The setup is thus like that of a one-to-one $S$-matrix, $p\rightarrow q$, in the background of $N$ particles with momenta $p_i$, see Fig.~\ref{SQFT}.

Finding  explicit examples in which the chaos effect in the $S$-matrix can be computed will be a challenge. Working perturbatively to a few orders in the coupling is not sufficient. Moreover, replacing the background state of $N$ particles $p_i$ with a thermal state is not appropriate; this would wash out the chaos \cite{Note4}

Of course, since relativistic quantum field theory subsumes non-relativistic quantum mechanics, with appropriate choices of initial state one can achieve, to arbitrarily good accuracy, scattering in classical mechanics or in quantum mechanics. For instance, one could take an $in$ state that gives  a long-lived intermediate state that looks like three disks. It would be interesting to study scattering of a relativistic particle in this background, thereby extending the non-relativistic analysis. 
One would hope, however, to also be able to see chaos in the $S$-matrix  in regions of parameter space in which there is no  intermediate classical state.

\subsubsection*{Chaos in the black hole $S$-matrix}
\begin{figure}[b]
\centering
\includegraphics[width=1in]{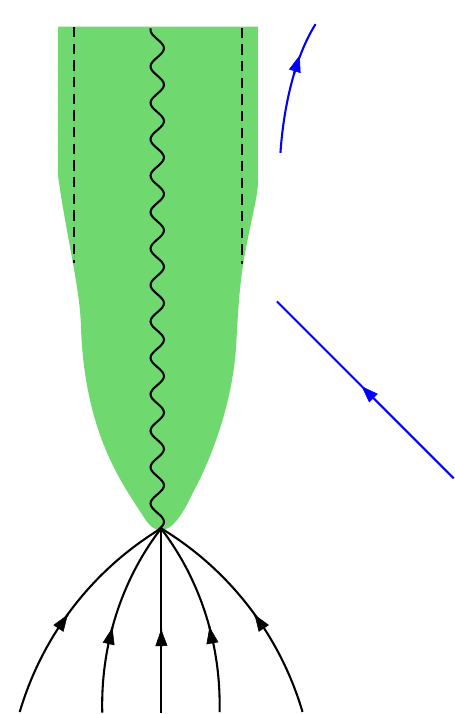}
\caption{ A sector of the quantum gravity $S$-matrix which exhibits chaos \cite{Polchinski}. Incoming particles collide to form a black hole as an intermediate state. The green shaded region is the interior of the black hole;   the edge  is the horizon. An additional particle falls into the black hole (blue line on the  right). The dashed vertical  line indicates where the horizon would have been had the extra particle not been sent in. The outgoing Hawking quanta, one of which is shown, are closer to the new horizon than to the old horizon.}\label{BH}
\end{figure}
There is one case in which chaos in a field theory $S$-matrix has in a sense already been seen, in the context of a black hole, as discussed in   \cite{Polchinski}, building on earlier work \cite{Dray:1984ha, tHooft:1990fkf, Shenker:2013pqa, Kitaev}. By appropriately sending in a large number of particles, we can form a black hole as an intermediate state, which decays into Hawking quanta at late times. The question is what impact a change in one of the $in$ particles has on the $S$-matrix. 
The setup, shown in Fig.~\ref{BH}, is the following: we add an additional particle to the $in$ state, so that the $in$ state has $N{+}1$ particles, where the additional particle interacts with the others much later, at a time at which they have already formed a black hole that is decaying into Hawking radiation. The effect of the additional particle is to shift the black hole horizon slightly outward. The subsequent outgoing Hawking quanta find themselves slightly closer to the horizon, which in turn causes them to take longer to escape.
More precisely, in Schwarzschild coordinates,  and for a horizon of radius   $r_s$, a Hawking quanta that is at radius $r_s+\delta$ takes a Schwarzschild time $t\sim \frac{\beta}{2\pi} \log \frac{r_s}{\delta}$ to escape. Decreasing $\delta\rightarrow \delta - \Delta r_s$ increases the escape time by an amount $\Delta t \sim \Delta r_s\, e^{2 \pi t/\beta}$, where $\beta$ is the inverse temperature of the black hole, which is proportional to its mass, which we can take to be the center of mass energy of the $in$ state \cite{Note2}.
This is chaos:  slightly increasing the energy of the $in$ state causes an exponentially large change in the $out$ state. \\[5pt]

The Horowitz-Polchinski correspondence principle \cite{HorowitzPolchinski} relates black holes and strings, suggesting one may be able to observe chaos in scattering involving highly excited strings. This is indeed the case and will be discussed for bosonic string theory in \cite{GRhigh}.
An excited string is characterized by the occupation numbers of the different modes, and can be formed by repeatedly scattering photons  off of an initial unexcited string (a tachyon), with the occupation numbers determining the momenta of the photons. Consequently, scattering of highly excited strings is an example of an $S$-matrix with a large number of particles, and the observed erratic behavior of the $S$-matrix in  \cite{GRhigh} is an illustration of the chaos effect proposed here.

\section*{Discussion}
Decades of progress in quantum field theory have given rise to both special field theories that are tractable even at finite coupling, as well as quantitative statements - such as a $c$-theorems - that are valid for all theories. Nevertheless,  even  a qualitative understanding of the dynamics of general quantum field theories remains an open problem. The primacy of the $S$-matrix within quantum field theory, combined with the generic appearance of chaos in classical field theory, suggests that studying chaos in the $S$-matrix of quantum field theory  may be a productive route. We have proposed one look for erratic behavior of the many-particle $S$-matrix under a small change in the momenta of a few particles. It will be important to find explicit examples of this, and to give a quantitative measure of the degree of chaos.

\section*{Acknowledgements} \noindent I am grateful to  U.~Smilansky for discussions on chaotic scattering and  the three-disk example, and to  B.~Shraiman for discussions on the leaky torus.  I  thank C.~Cordova, D.~Gross, A.~Kitaev, and M.~Smolkin for helpful discussions. 
This work was supported  by NSF grant PHY-1911298 and the Sivian Fund of the IAS..

\bibliographystyle{utphys}

\end{document}